\documentclass[conference]{IEEEtran}
\IEEEoverridecommandlockouts
\usepackage[square,sort,comma,numbers]{natbib}
\usepackage{amsmath,amssymb,amsfonts}
\usepackage{algorithmic}
\usepackage{graphicx}
\usepackage{textcomp}
\usepackage{xcolor}
\def\BibTeX{{\rm B\kern-.05em{\sc i\kern-.025em b}\kern-.08em
    T\kern-.1667em\lower.7ex\hbox{E}\kern-.125emX}}
    
\usepackage{hyperref}
\usepackage{setspace}
\usepackage{csquotes}
\usepackage{soul}

\begin{document}

\title{On Artificial Life and Emergent Computation in Physical Substrates\\
\thanks{This work was conducted as part of the SOCRATES project, which is partially funded by the Norwegian Research Council (NFR) through their IKTPLUSS research and innovation action on information and communication technologies under the project agreement 270961.}
}

\author{\IEEEauthorblockN{1\textsuperscript{st} Kristine Heiney}
\IEEEauthorblockA{\textit{Department of Computer Science} \\
\textit{Oslo Metropolitan University}\\
Oslo, Norway \\
kristine.heiney@oslomet.no}
\and
\IEEEauthorblockN{2\textsuperscript{nd} Gunnar Tufte}
\IEEEauthorblockA{\textit{Department of Computer Science} \\
\textit{Norwegian University of Science and Technology}\\
Trondheim, Norway \\
gunnar.tufte@ntnu.no}
\and
\IEEEauthorblockN{3\textsuperscript{rd} Stefano Nichele}
\IEEEauthorblockA{\textit{Department of Computer Science} \\
\textit{Oslo Metropolitan University,}\\
\textit{Department of Holistic Systems}\\
\textit{Simula Metropolitan}\\
Oslo, Noway \\
stenic@oslomet.no}
}

\maketitle

\begin{abstract}
In living systems, we often see the emergence of the ingredients necessary for computation---the capacity for information transmission, storage, and modification---begging the question of how we may exploit or imitate such biological systems in unconventional computing applications.
What can we gain from artificial life in the advancement of computing technology?
Artificial life provides us with powerful tools for understanding the dynamic behavior of biological systems and capturing this behavior in manmade substrates.
With this approach, we can move towards a new computing paradigm concerned with harnessing emergent computation in physical substrates not governed by the constraints of Moore's law and ultimately realize massively parallel and distributed computing technology.
In this paper, we argue that the lens of artificial life offers valuable perspectives for the advancement of high-performance computing technology.
We first present a brief foundational background on artificial life and some relevant tools that may be applicable to unconventional computing.
Two specific substrates are then discussed in detail: biological neurons and ensembles of nanomagnets.
These substrates are the focus of the authors' ongoing work, and they are illustrative of the two sides of the approach outlined here---the close study of living systems and the construction of artificial systems to produce life-like behaviors.
We conclude with a philosophical discussion on what we can learn from approaching computation with the curiosity inherent to the study of artificial life.
The main contribution of this paper is to present the great potential of using artificial life methodologies to uncover and harness the inherent computational power of physical substrates toward applications in unconventional high-performance computing.
\end{abstract}


\section{Introduction}
The field of artificial life concerns itself with how to produce complex macroscopic behaviors from the interaction of many simple interacting components. Where biology works to understand existing organisms and the complicated machineries that underlie observed physiological behaviors using an approach of deconstruction and element-by-element description, artificial life seeks to construct systems displaying interesting emergent behaviors by aggregating many simple objects governed by basic rules \cite{Langton1987}. Here, ``emergent'' refers to some feature of the entire system that cannot be described by the constituent parts of the system. For example, the physical concept of pressure has no meaning when considering only one or a few individual gas molecules; it is only when a large volume of gas is considered that this characteristic emerges as a meaningful descriptor of the system. Similarly, the movement and behavior of a single ant is qualitatively distinct from that of an entire colony. In many ways, these emergent behaviors may be seen as a form of computation, with the systems or organisms providing the machinery by which computations are performed.


The tools of artificial life allow us to capture these emergent behaviors without explicitly encoding them into the system. Rather, by creating simple sets of rules to describe the behavior of individual agents within the system as they move, connect, and interact, complex behaviors emerge of their own accord.
This approach to modeling and engineering dynamical systems can offer new perspectives on how to perform computation in substrates showing emergent properties, positioning us to answer the question posed by Langton \cite{Langton1990} for targeted physical systems: under what conditions might the capacity to perform computation emerge in a physical system?
Ongoing work being conducted by the authors involves the study of two physical substrates---networks of biological neurons \cite{Aaser2017TowardsMA, Heiney2019} and nanomagnetic ensembles \cite{Jensen2018}---along with models of these systems at different levels of abstraction \cite{PontesFilho2020}, and we argue that this inquisitive and bottom-up approach to understanding, mimicking, and constructing dynamical systems will prove fruitful in the advancement of bio-inspired parallel and distributed computing technology.

Sipper \cite{Sipper1999} highlights the three cornerstones of this computational paradigm, which he terms ``cellular computing'': simplicity, vast parallelism, and locality. In this paradigm, computation is performed with a vast number of very simple fundamental units whose connections are sparse and most often in the immediate vicinity. Thanks to this local connectivity, these machines thus perform without any centralized control, and their function is resilient against faults in the system.
Currently, much exploration into cellular computing is confined to simulation, though the ultimate aim is to construct actual machines, be they biological \cite{Benenson2012, Heiney2019} or manmade \cite{Jensen2018}, that can realize this behavior.
In comparison with classic von Neumann computing technologies, machines based in cellular computation principles will be more scalable, energy efficient, and resilient to failures of single elements \cite{Sipper1999,Das2016}.

This paper first explores basic concepts and motivations in artificial life and cellular computing.
A selection of approaches to capturing in models components of biological processes we see in nature are then presented.
Finally, the two abovementioned physical substrates---neuronal networks and nanomagnetic arrays---are explored in greater detail, and a philosophical discussion on how the lens of artificial life may inform our investigation of these substrates is presented.

\section{An Artificial Life Perspective on Complexity}
Human curiosity has long driven us to uncover how organisms function---what drives their behaviors on a macroscopic scale and what microscopic processes control physiological function. This has in turn driven many to explore means to recreate lifelike behaviors using mechanical components---these are the machineries of artificial life.

\begin{figure}[t]
\begin{center}
\includegraphics[width=0.75\linewidth]{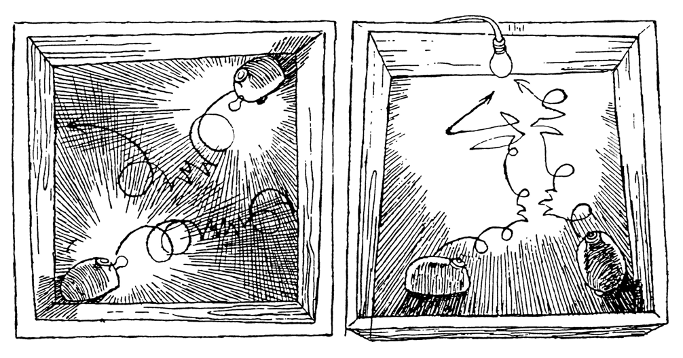}
\caption{Schematic of Grey Walter’s electronic tortoises, Elsie and Elmer. Reproduced from Walter \cite{walter1950}.}
\label{fig1tortoises}
\end{center}
\end{figure}

Early efforts in the field of artificial life focused on imitating behaviors observed in natural systems. For example, in 1950, William Grey Walter built a pair of artificial electronic ``tortoises'' named Elsie and Elmer that moved toward dim light but away from bright light \cite[Fig.\ \ref{fig1tortoises};][]{walter1950}. When he attached lights to the tortoises themselves, their interaction resulted in complex and interesting behavior, which he described as giving ``an eerie expression of purposefulness, independence and spontaneity'' \cite{walter1950}. Grey Walter took great delight in learning from these tortoises, named for the Mock Turtle's teacher in \textit{Alice's Adventures in Wonderland}, who, aptly, was also not exactly a tortoise: when Alice asked about the choice of name, the Mock Turtle replied,
``We called him Tortoise because he taught us.'' Although Grey Walter also predicted our captivation with the ``marvelous processes'' of life may ebb as our aptitude to imitate it grows, that prediction has not borne itself to fruition. Rather, it seems that the better able we are to capture complex dynamics in artificial systems, the greater our appreciation for the natural systems we strive to understand and emulate.

But what makes a system ``artificial''? It is not simply the mechanics or the fact that it is inorganic, nor is it the behavior the system shows, as this behavior is meant to be as close to that of the natural systems that inspire its construction. Herbert Simon gave an elegant definition of the artificial in his book The Sciences of the Artificial \cite{Simon1996}:
\vspace{-1mm}
\begin{displayquote}
 ``Artificiality connotes perceptual similarity but essential difference, resemblance from without rather than within… [W]e may say that the artificial object imitates the real by turning the same face to the outer system, by adapting, relative to the same goals, to comparable ranges of external tasks.''
\end{displayquote}
\vspace{-1mm}
Simon’s discussion in this section centers on the distinction between the task fulfilled by a designed system and the capability of the system itself, and what can be accomplished by artificial and simulated systems. We may create artificial systems that are perceptually similar to natural systems, systems that produce precisely the behaviors we wish to see on the scale at which we wish to see them, despite being inexorably different from within; this, indeed, is the situation we strive for in the study of artificial life.

What then do we do with such systems once we have captured their behavior? In some cases, the developed systems represent a means to understanding the system they mimic; in others, the system is the final product and gives us some capability that would be otherwise unattainable in conventional man-made systems. Models and simulations may provide insight into systems that would be unattainable by mere inspection of the assumptions and laws employed to govern the simulated system. This is useful when the precise mechanisms governing the behavior of a system are known at a certain scale but the system dynamics become more difficult to describe when the system is scaled up by the addition of more such simple components. Along the same lines, this complex larger-scale behavior can be useful in decoding the response of the system to different types of inputs, enabling the use of a dynamical system as a computational system.

However, the pursuits of researchers interested in artificial life have often been motivated not by some end goal but simply by curiosity. Much of the burgeoning research that has recently been conducted in the realm of artificial intelligence has been preoccupied with building better classifiers, better computer vision, better autonomous systems---summarily, the optimization of tools used as a means for computational tasks. Artificial life research is not necessarily oriented to these aims but serves to explore how we can harness the dynamics of behaviors of interest we observe in nature. As Langton put it in his paper on artificial life, artificial intelligence ``has focused primarily on the production of intelligent solutions rather than on the production of intelligent behavior. There is a world of difference between these two possible foci'' \cite{Langton1987}. Thus, in line with this curious, process-driven approach, this paper does not consider in great detail what we hope to ultimately do with the behaviors we observe as a motivation for studying artificial life but instead engages with the behaviors themselves as a point of interest with the assumption that applications and advancements will fall out as a natural product of the knowledge we gain with the approaches described here.

\section{Artificial Life Models and Tools for Unconventional Computing}
When considering the use of complex systems for computation, a system should be capable of three basic operations \cite{Langton1990}: the transmission, storage, and modification of information. To find systems that support these behaviors, we must be able to construct models of complex systems, develop metrics to quantify their performance, and search the space of all possible models to target the desired behaviors.

This section presents a general approach to capturing dynamic systems in models and characterizing their behavior. The concepts presented here represent some of the tools that are commonly used to explore the smaller-scale mechanisms underlying larger-scale complex dynamics. Dynamic systems are often modeled using an approach in which the system is considered to comprise a number of discrete components that interact with a set of predefined rules. Each of these components can be in a given state and can influence or be influenced by the states of a number of other components to which it is connected. Some examples of such systems---graphs and cellular automata (CAs)---will be presented in the following subsections.

The manner in which these dynamic models behave is governed by a set of parameters and rules defining the connections in the system, the states each component can take on, and how the state of each component influences the states of the components to which it is connected. If we wish to find the set of parameters and rules that produces a desired behavior in the system, we have a vast space in which to search, and that space is often largely occupied by many ``boring'' or undesirable models that fail to produce the behavior we seek. Thus, algorithms have been developed based on the biological principle of evolution to allow for a more targeted searching of the space of possibilities in model design. This approach will be discussed in more detail at the end of this section.

\subsection{Complex system models}
Complex systems are generally modeled using ensembles of discrete elements that interact with each other using a given set of rules. These elements can take on discrete or continuous states and can be connected to each other in a regular, irregular, or random fashion, or they can be agents allowed to move freely in space. This section presents two illustrative examples of different types of models used to represent the dynamical behavior of complex systems: graphs and CAs.

These two types of models were selected for their relevance to the question of computing in physical substrates. Graphs represent the connections between elements in a system and how those connections mediate the dynamics of the system; this type of model is very relevant in the field of neuroscience \cite{Sporns2000} and can capture the intricate arrangements of individual cells in a network or connections among brain regions. CAs, on the other hand, are one of the simplest representations of a dynamical system, and they offer insight into the vast range of dynamical behavior that can be achieved by tuning the parameters of the system, even in the case of a very simple elementary CA.

\subsubsection{Graphs}
A graph is composed of nodes, representing the discrete components of the system, and edges, representing the connections between the nodes. The state of each node affects the states of all of its neighbors in a manner defined by a set of rules or equations, much like the flight of a bird is affected by its fellow flock members. Graphs may also be constructed to have static or dynamic structures, with connections remaining fixed or evolving over time. The variation in the states of the nodes is referred to as dynamics on the network, whereas the change in the connections between the nodes is referred to as dynamics of the network. Systems that show both types of dynamics are referred to as dynamic systems with dynamic structures ((DS)$^2$) or adaptive networks. The brain is a prime example of an adaptive network, with numerous plasticity mechanisms constantly dynamically tweaking the weights of neural connections.

\begin{figure}[t]
\begin{center}
\includegraphics[width=0.85\linewidth]{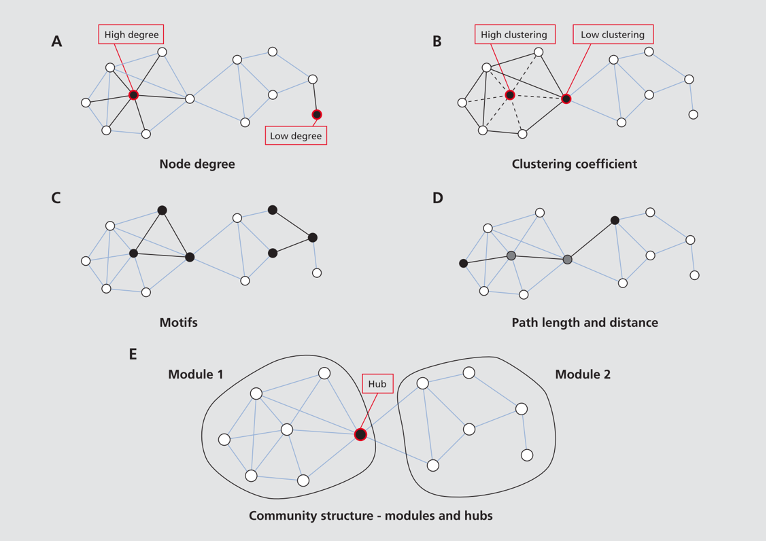}
\caption{Examples of graph theoretical measures commonly used in the study of complex systems. Reproduced from Sporns et al.~\cite{Sporns2013}.}
\label{fig2graphs}
\end{center}
\end{figure}

A number of useful measures can be extracted from graph maps of a given structure to give some insight into the dynamics on the network. Examples highlighted by Sporns et al.~\cite{Sporns2013} as applicable in the study of brain connectivity are shown in Fig.\ \ref{fig2graphs}, and further details on graph theoretical measures can be found in Sporns et al.~\cite{Sporns2000}.

Two network structures that represent opposite extremes are regular and random networks. Regular networks consist of nodes with the same number of neighbors and range from strongly regular, where every two adjacent nodes have the same number $n_1$ of neighbors in common and every two non-adjacent nodes have the same number $n_2$ of neighbors in common, to randomly regular, where every node has the same degree but the connections are randomly distributed. At the other extreme, random networks have a binomial degree distribution (or Poisson in the limit of a large number of nodes), meaning the nodes can be well-described by the average degree.

Many real systems tend to not show regular or random connectivity but lie somewhere in between these two extremes. Two types of commonly discussed models representative of real-world behaviors are scale-free and small-world networks. Scale-free networks have degree distributions that follow a power law: $P(k) \propto k^{-\gamma}$, where $P(k)$ is the fraction of nodes with degree $k$ and $\gamma$ is a constant. This means that many nodes have a low degree and few nodes have a high degree. Although many real networks have been reported to show scale-free connectivity, it is notoriously difficult to rigorously confirm power-law scaling from empirical data of finite systems \cite{Clauset2009}.

Scale-free networks are characterized by the presence of hubs and modularity. They tend to show local clustering and long-range integration and are robust against random removal or failure of nodes, as the vast majority of the nodes have few connections. However, if the hubs in the network are targeted, the system breaks down \cite[see, e.g.,][]{DelFerraro2018}. In comparison to a random network of the same size and average degree, the mean path length of a scale-free network is smaller whereas the clustering coefficient is much larger, demonstrating the modularity of the structure lends itself to efficient network communication.

Graph-based simulations can give insight into the dynamics of a number of different types of systems and how the interactions of their components can produce emergent macroscale behavior. Crucially, information flow through a system of interconnected elements can be readily represented with this modeling approach.

\subsubsection{Cellular automata}
A CA is classically defined as a regular $n$-dimensional lattice structure (or regularly connected graph) composed of discrete elements called cells that can take on discrete states. The state of each cell in the network progresses in discrete time steps according to a lookup table of rules that give the state at time step $t + 1$ based on the states of the cells in the neighborhood at time step $t$. Although CAs are actually a type of graph, the extra simplifying constraints placed on them make them a useful case to consider in the realm of computation.

The binary ($K=2$) one-dimensional CA with a neighborhood of size $N=3$ is one of the simplest possible dynamical system models to show complex behavior, and the range of behaviors that can be achieved by this model has been investigated in great depth. Such a model is an excellent exemplar of a complex system, as it shows a wide range of dynamic behavior that can serve as an analogue for the behavior of more complicated systems. In his seminal paper on the dynamical behavior of simple CAs, Langton \cite{Langton1990} explored the different behaviors that are attainable with this type of system and focused on how physical systems may show an emergent capacity for computation. This section will briefly explain the aims and achievements of his study.

\begin{figure}[t]
\begin{center}
\includegraphics[width=0.8\linewidth]{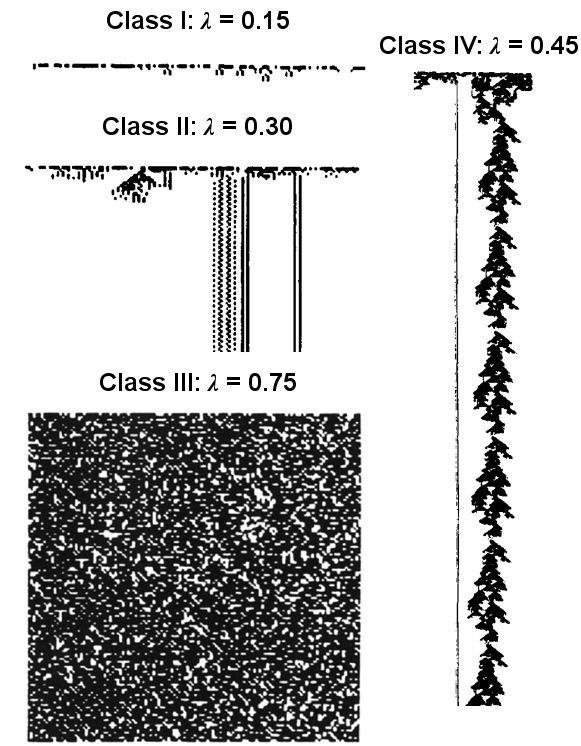}
\caption{Examples of CAs from the four different classes with $K=4$ states and $N=5$ neighbors. Images reproduced from \cite{Langton1990}.}
\label{fig4langton}
\end{center}
\end{figure}

As Langton \cite{Langton1990} stated, the focus of his paper was to determine ``the conditions under which [the] \textit{capacity to support computation} itself might emerge in physical systems'' by considering CAs as an exemplary simple model system. To this end, he qualitatively and quantitatively characterized a number of one-dimensional CAs with different rules and developed a new quantitative measure, the $\lambda$ parameter, that can be used to identify the qualitative class of behavior of a CA. This parameter is defined such that the cases where $\lambda=0.0$ and $1.0$ correspond to the most homogeneous and heterogeneous rulesets, respectively.

Wolfram \cite{Wolfram1984} had earlier defined four classes of CA behavior, with classes I and II corresponding to fixed and periodic behavior, class III showing aperiodic patterns with no identifiable structure, analogous to chaotic behavior.
Finally, class IV CAs yield ``complex patterns of localized structures,'' effectively having ``very long transients'' \cite{Langton1990}. These types of CAs show rich and interesting patterns of behavior with complex fractal-like structures emerging and propagating over space and time.
Langton’s survey of the possible rulesets for systems with $K=4$ and $N=5$ revealed an interesting correspondence between the qualitative behavior observed and the $\lambda$ parameter. Some examples of the observed behavior and the relationship between $\lambda$ and the class of behavior are shown in Fig.\ \ref{fig4langton}.

Shifting his focus to a larger two-dimensional CA, Langton \cite{Langton1990} also explored the relationships among $\lambda$, the average single-cell entropy $H$, and the mutual information (MI) between a cell and itself at the next time step. The relationships between these parameters revealed the presence of a sharp phase transition, corresponding to the transition between classes II and III. Langton’s results reveal high clustering of many CA rulesets in two distinct regions corresponding to these classes, with classes I and II occupying a region of low $H$ and low MI and class III a region of high $H$ and low MI. However, in the wide gap between these two regions lie a few sparse points representing the class IV CAs in the transitional regime poised delicately between order and chaos; here at a point of intermediate entropy, the MI is maximized in a sharp peak between the low-MI regions on either side.

Langton’s delving into the dynamics at the ``edge of chaos'' is a remarkably---and somewhat overwhelmingly---thorough exploration of how dynamics at the phase transitional regime may give insight into the nature of computation in the physical world. But what is most crucial to take away here? First, there is the more conceptual and more challenging lesson: computation can be accomplished by striking a precarious balance between information storage, which requires a lowering of the entropy, and information transmission, which requires raising the entropy. This lesson is applicable to any dynamical system, not only CAs. At the transitional regime between periodic and chaotic dynamics, we have the behavior needed for long-term and long-range correlations, allowing information to propagate arbitrarily far and remain for arbitrarily long periods of time.

Second, Langton’s parameter $\lambda$ gives us a way to survey the vast space of all possibilities to hone in on the systems that can show the behaviors we wish to see in computational systems. Considering how rapidly the number of possible systems can expand as a modeling framework is adjusted to represent ever more nuanced features of actual physical systems, this surveying ability is highly valuable. The following section will also address a more general approach to surveying the space of possible systems to find desired behavior.

\subsection{Evolutionary Algorithms}
As may have been apparent from the explanations of the two example model systems in the previous section, even relatively small systems with simple rules governing their local dynamics can often show very complex behavior that cannot be predicted by examining their structure and rulesets alone; rather, the system must be run to observe its behavior. For any interesting system, there will certainly exist a vast number of possible configurations and rulesets, to the extent that it would not be reasonable to brute force our way through checking the behavior of each one. Furthermore, as discussed by Langton \cite{Langton1990} and exemplified by his CA survey, the number of configurations showing interesting behavior becomes vanishingly small with respect to the space of all possible configurations as the system size scales up. All of these factors make it challenging to select for systems that show a targeted type of behavior.

One approach to tackling this issue is the use of evolutionary algorithms, which take inspiration from the process of evolution by natural selection to iteratively improve generations of machines to produce the desired behavioral outcome \cite{Langton1987}. In this approach, the rules that govern local interactions are encoded into a simple representation of a possible system configuration, and the behavior that is produced when the system is run is evaluated to determine how well it performs based on a desired metric called the fitness function. The representation and output behavior are conceptually similar to the genotype (genetic makeup) and phenotype (observable characteristics) of an organism.

The process of computational evolution then follows a path analogous to that in nature. An initial population of individual machines is created and run. Their fitness is evaluated based on a fitness function quantifying how closely their behavior resembles the target behavior. The descriptions of those that perform the best from the population (parent machines) are then used to generate new descriptions for a new generation of machines (offspring machines). For this purpose, genetic algorithms are commonly used. These involve selecting pairs of machines from the parent generation that show high fitness and performing genetics-inspired operations like crossover and mutation. This process is iterated over many generations and allows for a more intelligent search of the space of possible machines. In the realm of computation in physical substrates, a technique known as evolution-in-materio is commonly used \cite{Broersma2016}, where genetic algorithms are used to search for physical signals that can be applied to a physical system to configure its properties to a desired state, heightening its capacity for computation.

\section{Working with Physical Substrates}
To better understand the behaviors driving complex dynamics in actual physical systems, data-driven modeling approaches can be employed, where data obtained from actual physical substrates is obtained and analysis of this data is used to recapture targeted behaviors in models. In the case of engineerable substrates, targeted behaviors may even be translatable from natural biological substrates. Data-driven modeling serves a dual purpose: (1) providing insight into the behavior of the studied system, including through simulations of situations that may not be easily achievable experimentally, and (2) enabling the emulation of natural systems in artificial systems.

Furthermore, increasing attention has turned to the exploitation of physical nonlinear systems for computation \cite{Stepney2008}. Current computing substrates are inflexible and power-hungry; the use of complex nonlinear systems in computing hardware would open up the possibility for more efficient and powerful hardware with the capacity to learn and could pave the way for rapid advancement in artificial intelligence.
One example of a computing paradigm exploiting the nonlinearity of certain systems is reservoir computing \cite{Tanaka, LUKOSEVICIUS2009127}, one great benefit of which is that the system acting as the reservoir does not need to be trained or modified; rather, the connections and dynamical behaviors it shows can be harnessed by finding the appropriate way to encode inputs to apply to the system and decode the output behavior it produces.
This computing paradigm has been exploited in artificial intelligence applications \cite{LUKOSEVICIUS2009127}, and artificial life approaches to computational reservoirs may offer tools for the advancement of such applications.

This section gives an overview of the dynamics of two physical substrates  investigated by the authors---networks of neurons and nanomagnetic arrays---and some approaches that have been taken to capture their behavior in computational models. These two substrates both show the necessary nonlinear behavior for them to be well-suited for computation.

\subsection{Neuronal networks}
The human brain is arguably the most complicated machine we know of. With very low power consumption, it can make sense of an impressive range of inputs and control widely varied bodily responses to these inputs, and it can store vast amounts of information. Neurons in the brain encode and transmit information in stereotyped electrical signals called action potentials, or spikes, which are produced by a carefully synchronized flow of ions across the cell membrane and can induce spikes in other neurons via connections called synapses. Although the mechanisms of spike generation, propagation, and transmission are fairly well characterized at the cellular level---though, admittedly, even on a single-cell level, what is known is far from simple and there remains much to be learned---larger-scale behaviors cannot be wholly explained by simply combining many of these smaller-scale models.

Much research has been devoted to understanding the immense computational capabilities of the brain and how networks of neurons process, transmit, and store information, and advances in recording technology and data handling techniques have opened the door to new lines of investigation that were previously inaccessible. The electrical behavior of neurons allows them to be studied at many scales using different techniques, including single-cell recording by measuring the voltage across the cell membrane and electroencephalogram (EEG) at the whole-brain scale. To limit the scope of the discussion here, we focus on the study of the electrophysiological behavior of neurons at the network or population level using microelectrode array (MEA) technology \cite{Obien2015}.

\begin{figure}[t]
\begin{center}
\includegraphics[width=\linewidth]{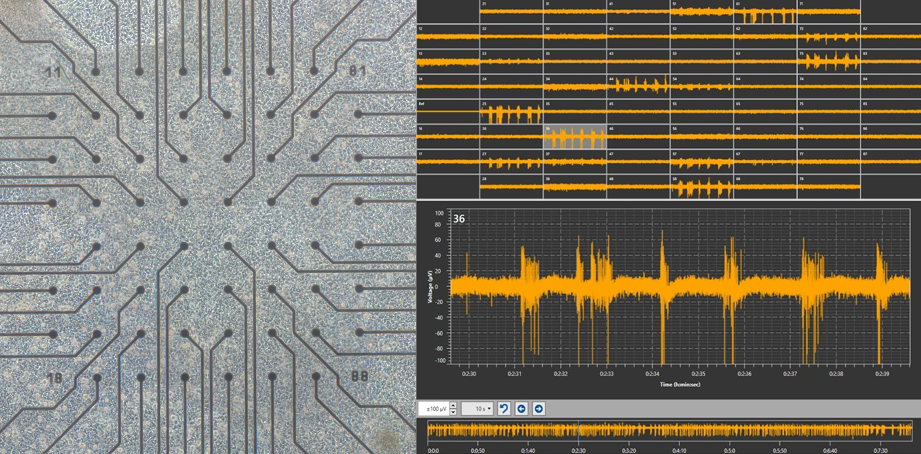}
\caption{Microscope image of an MEA (Multi Channel Systems GmbH, Germany) with a network of neurons cultured on top of it \cite{Heiney2019} (left) along with a screen capture from the Multi Channel Suite of software (Multi Channel Systems GmbH, Germany).}
\label{fig5neurons}
\end{center}
\end{figure}

An MEA is a set of electrodes embedded in a substrate, such as glass, on which neurons can be grown. MEAs enable the long-term nondestructive recording of populations of neurons as well as controlled network stimulation by electrical pulses. An example of a 60-electrode MEA is shown in Fig.\ \ref{fig5neurons}, along with an example of the voltage signals recorded by the MEA, from which the spiking behavior of the network can be extracted.

The advent and recent advancement of MEA technology, including accessibility to commercial recording setups and analysis tools (e.g., Multi Channel Systems MEA2100 systems and MEAs, along with their corresponding software suite; see Fig.\ \ref{fig5neurons}) have made it possible for researchers to perform long-term observation of the behavior of populations of neurons in vitro. This capacity to record from whole populations of neurons, both in living animals and in disembodied cultures, using MEAs and related technology has brought about a shift in focus from the spiking of single neurons to network-level dynamics and how neuronal assemblies within the brain can collectively drive the dynamics and function of the entire system \cite{Poli2015}. Indeed, although complex and inarguably worthy of the attention it has received, the behavior of a single neuron can only tell us so much when decoupled from its ``neighborhood'' of other interconnected neurons.

This shift in focus has brought about the necessity to inquire into the organization of such populations of neurons, raising a number of questions. What physical feature of the network constitutes a connection between two neurons? How can we capture these connections in the network through our observations of it? And what measures can we use to characterize the network connections?

\subsubsection{Connectivity in neuronal networks}
The organization of networks of neurons is typically described in terms of three types of connectivity: structural, functional, and effective connectivity \cite{Poli2015}. A structural connection indicates an anatomical feature that mediates a physical interaction between two neurons, namely a synapse. Structural connectivity is extracted from imaging data of morphological features or synaptic markers. Although methods exist for extracting such features in relatively low-density networks \cite[e.g.,][]{Ullo2014, MorenoeMello2019}, where the morphology of individual neurons can be observed, in higher-density networks on MEAs, such features can be difficult to extract from images.

Functional connectivity represents the temporal correlation between the spiking patterns of pairs of neurons obtained using, for example, the cross-correlation between pairs of spike trains. In this type of connectivity, two neurons are said to be connected if the spiking of one can be predicted from the spiking of the other; however, this does not necessarily indicate a causal relationship between the spiking of the two. The effective connectivity, in contrast, refers to connections where the activity of one neuron can be said to directly cause that of another neuron.

\begin{figure}[t]
\begin{center}
\includegraphics[width=\linewidth]{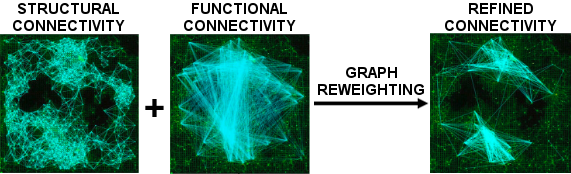}
\caption{Combined structural and functional connectivity to capture a more accurate representation of the organization and activity of an in vitro neuronal network. Adapted from Ullo et al.\ \cite{Ullo2014}.}
\label{fig6ullo}
\end{center}
\end{figure}

In one noteworthy study \cite{Ullo2014}, the structural and functional connectivity were obtained from imaging and electrophysiological data from a high-density MEA and then combined to yield a refined functional connectivity map that may be said to better indicate the actual organization and activity of the network (Fig.\ \ref{fig6ullo}). It should be noted that all of these types of connectivity are subject to change as a result of different types of plasticity mechanisms operating on a wide range of time scales. This plasticity means that neuronal systems show both dynamics on the network, in the form of spiking activity traveling through the network, and dynamics of the network, with the connections between neurons subject to changing as a result of their activity.

\subsubsection{Connectivity and dynamics: How does information flow?}
These types of connectivity give us an idea of what may be considered to constitute a connection and how to capture such connections---that is, either visual observation of an anatomical connection or temporal correlations in the electrophysiological spiking data recorded from pairs or neurons. What conclusions can then be drawn from these connectivity maps, and what tools can we use to get there?

A first clear step once the connectivity is obtained is to apply graph theory measures (see Fig.\ \ref{fig2graphs}) to characterize the organization of the connectivity map. These measures can tell us if, for example, the network is modular and contains many hubs or if it is more randomly connected with many nodes having roughly the same degree. The brain is known to strike a balance between functional segregation and functional integration, allowing information to flow between spatially distant parts of the brain while allowing the generation of coherent brain states \cite{Sporns2000}. Mapping the connectivity of neuronal populations may give us insight into how this balance is struck. Additionally, as mentioned previously, the connectivity of a network changes over time, both as a consequence of maturation and in response to stimuli. Connectivity analysis can provide us a deeper understanding of how neurons organize themselves over time and what may drive these organizational structures, in both healthy networks and networks that mimic diseases or other abnormal states, and can show how the connectivity is affected by the adaptive or maladaptive responses the networks have to external stimuli or perturbations.

The connectivity of a network is also intimately tied with the dynamics that happen on the network and the manner in which elements in the network communicate and process information. One approach to capturing the dynamic state of a network is to study the distribution of the size of network-wide cascades of activity called ``neuronal avalanches'' \cite{Beggs2003}. It has been theorized that the brain lies in the critical state, a state analogous to the ``edge of chaos'' explored by Langton \cite{Langton1990} in which information processing is optimized, and in this state, the size distribution of neuronal avalanches follows a power law. Massobrio et al.\ \cite{Massobrio2015} have shown that a model with scale-free connectivity is able to reproduce the power-law avalanche scaling we expect to see in networks at criticality, and Shew et al.\ \cite{Shew2009,Shew2011} have shown optimized dynamic range, information capacity, and information transmission in networks showing power-law avalanche scaling. Additionally, preliminary results indicate it may be possible to manipulate supercritical networks into the critical state by increasing network inhibition \cite{Heiney2019}, enabling comparative studies between their behavior in different states and their exploitation as computational reservoirs. These results demonstrate the link between connectivity, dynamic state, and information processing in neuronal networks.

Future work into these three perspectives on the dynamical behavior of neurons may give us invaluable insight into how we can construct self-organizing systems to show the same kind of capacity for efficient information processing---how we can organize connections between elements of the system and construct rules for how they affect each other’s behavior. Building models like this can in turn give us a deeper understanding of the brain as well, as we target specific behaviors to emulate and see how simplified models can produce behaviors analogous to those observed in the original system.

\subsection{Magnetic substrates for computation}
A number of substrates based in ensembles of different types of magnetic materials show interesting dynamic interplay between the elements arising from various physical phenomena, and these substrates are promising candidates for reservoir computing applications.
There are a number of recent examples of studies exploring the possibility of exploiting magnetic substrates for computation \cite[e.g.,][]{Bourianoff2018, Torrejon2017}; however, we focus here on a specific paper by Jensen et al.\ \cite{Jensen2018} concerned with exploring how to tune the dynamic behavior of artificial spin ice (ASI) by varying the parameters of an external driving field.
For a recent review of the use of physical substrates for reservoir computing, see Tanaka et al.\ \cite{Tanaka}.

\begin{figure}[t]
\begin{center}
\includegraphics[width=\linewidth]{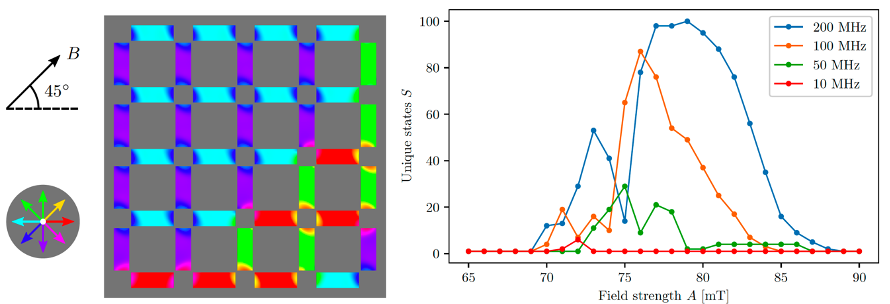}
\caption{Schematic showing the layout of the ASI studied by Jensen et al.\ \cite{Jensen2018} (left). The number of unique states visited by the array was counted for different external field parameters (right). Reproduced from \cite{Jensen2018}.}
\label{fig7magnets}
\end{center}
\end{figure}

ASI \cite{Skjervo2019} consists of an array of coupled nanomagnets arranged on a two-dimensional lattice. Each nanomagnet can be viewed as a single bit because of its dipolar state. The state behavior of these nanomagnetic islands arises from the small scale of their dimensions and their shape. Nanomagnets consist of a single such domain, meaning the magnetization does not vary across the magnet. Oftentimes, the nanomagnets will be fabricated with an elongated rectangular shape, causing there to be two states that are energetically favorable, termed spin up and spin down and corresponding to magnetization along the longitudinal axis in the positive or negative direction. This shape of nanomagnet was used in the square ASI array studied by Jensen et al.\ \cite{Jensen2018} (Fig.\ \ref{fig7magnets}).

Jensen et al.\ \cite{Jensen2018} perturbed this simulated nanomagnetic array with a time-varying external magnetic field, $B(t)=A \sin \omega t$, at a direction $45^\circ$ from the horizontal, and the field parameters $A$ and $\omega$ were tuned to achieve different types of dynamical behavior. 
Because each nanomagnet in the system can take on one of two states, the overall state of the system can be represented by 40 bits, yielding 240 possible states. To quantify the complexity of the behavior of the system, 100 cycles of the external field were applied to the system with different strengths $A$ and frequencies $\omega$, and the number $S$ of unique states observed at the end of every cycle was counted ($1\leq S \leq 100$). The results are shown in the right-hand panel of Fig.\ \ref{fig7magnets}.

For weak fields, none of the magnets switch their state ($S=1$), and for very strong fields, all of the magnets switch at the half-cycle point and then switch back ($S=1$ again). Additionally, at low frequencies, any transient behavior has died out by the end of a single cycle, so the number of states remains low, whereas very high frequencies produced chaotic behavior, with the nanomagnets not having sufficient time to ``keep up'' with the switching of the oscillating field direction. At the intermediate frequency of 100 MHz, long transients were observed in a manner analogous to the complex CA behavior seen in the right-hand panel of Fig.\ \ref{fig4langton}. 

This type of substrate cannot show the same kind of long-range connectivity that is observed in neuronal assemblies, as a single element of a magnetic system can only directly affect other elements within a certain local radius, whereas neurons can grow axons that can extend over very long distances within the network. Thus, the ``connectivity'' of ASI is governed by quite a different set of physical rules. However, the study described here demonstrates that complex behaviors can be captured in this substrate and information can propagate through the local interactions that occur between pairs of nanomagnets. By modeling the essence of the behaviors of neural systems that allow them to achieve the kinds of dynamics and optimal information processing behaviors described in the previous section, we can drive the development of magnetic substrates such as these towards a realm of greater efficiency and power, with the possibility of capturing some of the information processing capacity of the brain in engineerable computational substrates.

\section{Discussion}
The fields of complexity and artificial life offer a great many tools to study emergent behaviors and complex dynamics in different types of dynamical systems. Close inspection of such systems---be they model systems like cellular automata, natural systems like the brain, or fabricated systems like artificial spin ice---reveals rich and varied behaviors that are challenging to capture in simple metrics but beautiful to watch unfold.

With the use of models like graphs and CAs, we can capture the wide range of possible dynamical behaviors observed in various physical systems and target those behaviors that align with the hallmarks of computational power and edge-of-chaos dynamics.
This vanishingly small space of critical systems described by Langton \cite{Langton1990} and Wolfram \cite{Wolfram1984} is accessible to us if we know how or where to search, and tools like evolutionary algorithms put this possibility at our fingertips.
Furthermore, with the connection between neuronal avalanches and criticality \cite{Beggs2003}, we may also work backward from the product---the neuronal system identified as complex---to the description of its connectivity and response to inputs.
From this understanding our focus may then turn to eliciting from manmade substrates the same capacity for computation we see in natural systems.
However, it is important to remember that these models and neuro-inspired substrates \textit{are not} the brain, nor do they \textit{behave} precisely as the brain does---remember again Simon's (1996) statement: ``the artificial object ... turn[s] the same face to the outer system.''
Interpretations of the behavior we capture in our models must be tempered with this understanding: that as we mimic, we do not precisely recreate, and there may be just as much to gain from the differences as the similarities.

An artificial life approach can provide solutions to a wide range of practical questions.
The focus here has been on bio-inspired parallel and distributed computation, with potential applications spanning from the implementation of biologically plausible models for computation to the development of bio-inspired computational substrates with the ability to learn and adapt, offering a physical environment better suited for artificial intelligence applications than conventional hardware.
In addition, teaming up with biologists and neuroscientists to better characterize the dynamics of the brain may open the door to previously unconsidered diagnostic or clinical tools.

But apart from the practical, these systems spark in many researchers an innate and powerful curiosity needing no pragmatic outlet. We argue this drive to understand how complex behaviors emerge from tauntingly simple components and rulesets describing their interaction---coupled with a mind open to the possibilities of what an exploration of these systems will reveal---is what will ultimately prove fruitful in future research, giving us fodder for the practical applications where such answers were not at first sought. There still remains much for the tortoises to teach us.

\bibliographystyle{IEEEtran}
\bibliography{IEEEabrv,refs}

\begin{thebibliography}{10}
\providecommand{\url}[1]{#1}
\csname url@samestyle\endcsname
\providecommand{\newblock}{\relax}
\providecommand{\bibinfo}[2]{#2}
\providecommand{\BIBentrySTDinterwordspacing}{\spaceskip=0pt\relax}
\providecommand{\BIBentryALTinterwordstretchfactor}{4}
\providecommand{\BIBentryALTinterwordspacing}{\spaceskip=\fontdimen2\font plus
\BIBentryALTinterwordstretchfactor\fontdimen3\font minus
  \fontdimen4\font\relax}
\providecommand{\BIBforeignlanguage}[2]{{%
\expandafter\ifx\csname l@#1\endcsname\relax
\typeout{** WARNING: IEEEtran.bst: No hyphenation pattern has been}%
\typeout{** loaded for the language `#1'. Using the pattern for}%
\typeout{** the default language instead.}%
\else
\language=\csname l@#1\endcsname
\fi
#2}}
\providecommand{\BIBdecl}{\relax}
\BIBdecl

\bibitem{Langton1987}
C.~G. Langton, ``{Artificial Life},'' in \emph{Proceedings of the
  Interdisciplinary Workshop on the Synthesis and Simulation of Living
  Systems}.\hskip 1em plus 0.5em minus 0.4em\relax Addison-Wesley Publishing
  Company, 1987, pp. 1--48.

\bibitem{Langton1990}
C.~Langton, ``{Computation at the edge of chaos: Phase transitions and emergent
  computation},'' \emph{Physica D: Nonlinear Phenomena}, vol.~42, no. 1-3, pp.
  12--37, 1990.

\bibitem{Aaser2017TowardsMA}
P.~Aaser, M.~Knudsen, O.~H. Ramstad, R.~van~de Wijdeven, S.~Nichele,
  I.~Sandvig, G.~Tufte, U.~S. Bauer, {\O}.~Halaas, S.~Hendseth, A.~Sandvig, and
  V.~D. Valderhaug, ``Towards making a cyborg: A closed-loop reservoir-neuro
  system,'' in \emph{ECAL}, 2017.

\bibitem{Heiney2019}
K.~{Heiney}, O.~H. {Ramstad}, I.~{Sandvig}, A.~{Sandvig}, and S.~{Nichele},
  ``Assessment and manipulation of the computational capacity of in vitro
  neuronal networks through criticality in neuronal avalanches,'' in \emph{2019
  IEEE Symposium Series on Computational Intelligence (SSCI)}, 2019, pp.
  247--254.

\bibitem{Jensen2018}
J.~H. Jensen, E.~Folven, and G.~Tufte, ``{Computation in artificial spin
  ice},'' in \emph{The 2018 Conference on Artificial Life}.\hskip 1em plus
  0.5em minus 0.4em\relax Cambridge, MA: MIT Press, 2018, pp. 15--22.

\bibitem{PontesFilho2020}
S.~Pontes-Filho, P.~Lind, A.~Yazidi, J.~Zhang, H.~Hammer, G.~Mello, I.~Sandvig,
  G.~Tufte, and S.~Nichele, ``Evodynamic: A framework for the evolution of
  generally represented dynamical systems and its application to criticality,''
  in \emph{EvoApplications 2020, Held as part of EvoStar 2020}, 2020.

\bibitem{Sipper1999}
M.~Sipper, ``The emergence of cellular computing,'' \emph{Computer}, vol.~32,
  no.~7, pp. 18--26, 1999.

\bibitem{Benenson2012}
Y.~Benenson, ``Biomolecular computing systems: principles, progress and
  potential,'' \emph{Nature Reviews Genetics}, vol.~13, no.~7, pp. 455--468,
  2012.

\bibitem{Das2016}
A.~Das, R.~Dasgupta, and A.~Bagchi, ``Overview of cellular computing-basic
  principles and applications,'' in \emph{Handbook of Research on Natural
  Computing for Optimization Problems}, J.~K. Mandal, S.~Mukhopadhyay, and
  T.~Pal, Eds.\hskip 1em plus 0.5em minus 0.4em\relax Hershey, PA, USA: IGI
  Global, 2016, pp. 637--662.

\bibitem{walter1950}
W.~{Grey Walter}, ``An imitation of life,'' \emph{Scientific American}, pp.
  42--45, 1950.

\bibitem{Simon1996}
H.~A. Simon, \emph{{The Sciences of the Artificial}}, 3rd~ed.\hskip 1em plus
  0.5em minus 0.4em\relax Cambridge, MA: MIT Press, 1996.

\bibitem{Sporns2000}
O.~Sporns, G.~Tononi, and G.~M. Edelman, ``{Connectivity and complexity: The
  relationship between neuroanatomy and brain dynamics},'' \emph{Neural
  Networks}, vol.~13, no. 8-9, pp. 909--922, 2000.

\bibitem{Sporns2013}
O.~Sporns, ``{Structure and function of complex brain networks},''
  \emph{Dialogues in Clinical Neuroscience}, vol.~15, no.~3, pp. 247--62, 2013.

\bibitem{Clauset2009}
A.~Clauset, C.~R. Shalizi, and M.~E. Newman, ``{Power-law distributions in
  empirical data},'' \emph{SIAM Review}, vol.~51, no.~4, pp. 661--703, 2009.

\bibitem{DelFerraro2018}
G.~{Del Ferraro}, A.~Moreno, B.~Min, F.~Morone,
  {\'{U}}.~P{\'{e}}rez-Ram{\'{i}}rez, L.~P{\'{e}}rez-Cervera, L.~C. Parra,
  A.~Holodny, S.~Canals, and H.~A. Makse, ``{Finding influential nodes for
  integration in brain networks using optimal percolation theory},''
  \emph{Nature Communications}, vol.~9, no.~1, p. 2274, 2018.

\bibitem{Wolfram1984}
S.~Wolfram, ``{Universality and complexity in cellular automata},''
  \emph{Physica D: Nonlinear Phenomena}, vol.~10, pp. 1--35, 1984.

\bibitem{Broersma2016}
H.~J. Broersma, J.~F. Miller, and S.~Nichele, \emph{Computational matter:
  evolving computational functions in nanoscale materials}, ser. Emergence,
  Complexity and Computation.\hskip 1em plus 0.5em minus 0.4em\relax Springer,
  2016, no.~23, pp. 397--428.

\bibitem{Stepney2008}
S.~Stepney, ``{The neglected pillar of material computation},'' \emph{Physica
  D: Nonlinear Phenomena}, vol. 237, no.~9, pp. 1157--1164, 2008.

\bibitem{Tanaka}
G.~Tanaka, T.~Yamane, J.~B. Héroux, R.~Nakane, N.~Kanazawa, S.~Takeda,
  H.~Numata, D.~Nakano, and A.~Hirose, ``Recent advances in physical reservoir
  computing: A review,'' \emph{Neural Networks}, vol. 115, pp. 100--123, 2019.

\bibitem{LUKOSEVICIUS2009127}
M.~Luko{\v{s}}evi{\v{c}}ius and H.~Jaeger, ``{Reservoir computing approaches to
  recurrent neural network training},'' \emph{Computer Science Review}, vol.~3,
  no.~3, pp. 127--149, 2009.

\bibitem{Obien2015}
M.~E.~J. Obien, K.~Deligkaris, T.~Bullmann, and D.~J. Bakkum, ``{Revealing
  neuronal function through microelectrode array recordings},'' \emph{Frontiers
  in Neuroscience}, vol.~8, pp. 1--30, 2015.

\bibitem{Poli2015}
D.~Poli, V.~P. Pastore, and P.~Massobrio, ``{Functional connectivity in in
  vitro neuronal assemblies},'' \emph{Frontiers in Neural Circuits}, vol.~9,
  pp. 1--14, 2015.

\bibitem{Ullo2014}
S.~Ullo, T.~R. Nieus, D.~Sona, A.~Maccione, L.~Berdondini, and V.~Murino,
  ``{Functional connectivity estimation over large networks at cellular
  resolution based on electrophysiological recordings and structural prior},''
  \emph{Frontiers in Neuroanatomy}, vol.~8, pp. 1--15, 2014.

\bibitem{MorenoeMello2019}
G.~B. {Moreno e Mello}, S.~Pontes-Filho, I.~Sandvig, V.~D. Valderhaug,
  E.~Zouganeli, O.~H. Ramstad, A.~Sandvig, and S.~Nichele, ``{Method to Obtain
  Neuromorphic Reservoir Networks from Images of in Vitro Cortical Networks},''
  in \emph{2019 IEEE Symposium Series on Computational Intelligence
  (SSCI)}.\hskip 1em plus 0.5em minus 0.4em\relax IEEE, 2019, pp. 2360--2366.

\bibitem{Beggs2003}
J.~M. Beggs and D.~Plenz, ``{Neuronal Avalanches in Neocortical Circuits},''
  \emph{Journal of Neuroscience}, vol.~23, no.~35, pp. 11\,167--11\,177, 2003.

\bibitem{Massobrio2015}
P.~Massobrio, V.~Pasquale, and S.~Martinoia, ``{Self-organized criticality in
  cortical assemblies occurs in concurrent scale-free and small-world
  networks},'' \emph{Nature Publishing Group}, vol.~5, 2015.

\bibitem{Shew2009}
W.~L. Shew, H.~Yang, T.~Petermann, R.~Roy, and D.~Plenz, ``{Neuronal avalanches
  imply maximum dynamic range in cortical networks at criticality},''
  \emph{Journal of Neuroscience}, vol.~29, no.~49, pp. 15\,595--15\,600, 2009.

\bibitem{Shew2011}
W.~L. Shew, H.~Yang, S.~Yu, R.~Roy, and D.~Plenz, ``{Information capacity and
  transmission are maximized in balanced cortical networks with neuronal
  avalanches},'' \emph{Journal of Neuroscience}, vol.~31, no.~1, pp. 55--63,
  2011.

\bibitem{Bourianoff2018}
G.~Bourianoff, D.~Pinna, M.~Sitte, and K.~Everschor-Sitte, ``{Potential
  implementation of reservoir computing models based on magnetic skyrmions},''
  \emph{AIP Advances}, vol.~8, no.~5, p. 055602, 2018.

\bibitem{Torrejon2017}
J.~Torrejon, M.~Riou, F.~A. Araujo, S.~Tsunegi, G.~Khalsa, D.~Querlioz,
  P.~Bortolotti, V.~Cros, K.~Yakushiji, A.~Fukushima, H.~Kubota, S.~Yuasa,
  M.~D. Stiles, and J.~Grollier, ``{Neuromorphic computing with nanoscale
  spintronic oscillators},'' \emph{Nature}, vol. 547, no. 7664, pp. 428--431,
  2017.

\bibitem{Skjervo2019}
S.~H. Skj{\ae}rv{\o}, C.~H. Marrows, R.~L. Stamps, and L.~J. Heyderman,
  ``Advances in artificial spin ice,'' \emph{Nature Reviews Physics}, vol.~2,
  pp. 13--28, 2020.

\end{thebibliography}

\end{document}